\documentstyle[preprint,aps,prd]{revtex}
\def\be{\begin{equation}}
\def\ee{\end{equation}}
\begin{document}
\draft 
\title{Gravitational Lorentz Force and the Description \\ 
of the Gravitational Interaction}
\author{V. C. de Andrade and J. G. Pereira}
\vskip 0.5cm
\address{Instituto de F\'{\i}sica Te\'orica\\
Universidade Estadual Paulista\\
Rua Pamplona 145\\
01405-900\, S\~ao Paulo \\ 
Brazil}
\maketitle

\begin{abstract}

In the context of a gauge theory for the translation group, we
have obtained, for a spinless particle, a gravitational analog
of the Lorentz force. Then, we have shown that this force 
equation can be rewritten in terms of magnitudes related to either
the teleparallel or the riemannian structures induced in spacetime
by the presence of the gravitational field. In the first case, 
it gives a force equation, with torsion playing the role of force. 
In the second, it gives the usual geodesic equation of General 
Relativity. The main conclusion is that scalar matter is able to 
feel anyone of the above spacetime geometries, the teleparallel 
and the metric ones. Furthermore, both descriptions are found to 
be completely equivalent in the sense that they give the same 
physical trajectory for a spinless particle in a gravitational
field. 
\end{abstract}


\vfill \eject

\section{Introduction}

The notion of absolute parallelism (or teleparallelism) was 
introduced by Einstein in the late twenties, in his attempt to 
unify gravitation and electromagnetism. About three decades later, 
works by M{\o}ller~\cite{moller}, Pellegrini \&
Plebanski~\cite{pelle}, and  Hayashi \& Nakano~\cite{haya},
produced a revival of those ideas, which since then have received
considerable attention, mainly in the context of gauge theories
for the Poincar\'e and the translation
groups~\cite{hayshi,hehl,kopc,azeredo,nitsch1,mielke,hene}.

The scene of the teleparallel  theories of gravitation is the
Weitzenb\"ock spacetime~\cite{weitz}, a space presenting torsion,
but no curvature. The teleparallel description of gravitation is
believed to be equivalent, at least macroscopically,
to the General Relativity description, 
whose stage--set is provided by a Riemann spacetime,
a space presenting curvature, but no torsion. If this equivalence
is in fact true and effective, the gravitational interaction might
have two equivalent descriptions, one of them in terms of torsion
only, and another one in terms of curvature only.

With the purpose of exploring this equivalence, we study in this 
paper a gauge theory for the translation group, trying to stay as
close as possible to the usual scheme of the gauge models for internal 
groups. This means essentially that we start by considering spacetime 
to be a Minkowski space. The resulting model, as we are going to see,
will be quite analogous to the $U(1)$ electromagnetic gauge
theory. Relying on this analogy, and  considering the motion of a
spinless test particle in a translational gauge gravitational
field, we deduce the gravitational analog of the  Lorentz force
equation. This equation describes the trajectory of the particle
submitted to a gauge gravitational field in a flat spacetime. Now, 
due to the spacetime character of translations, the
corresponding gauge theory will differ from the usual gauge models
in many ways, the most significant being the presence of a tetrad 
field. A tetrad field defines in a natural way a linear Cartan
connection  with respect to which the tetrad is parallel. For this
reason, tetrad theories have received the name of teleparallelism,
or absolute parallelism. A tetrad field defines also in a natural
way a riemannian metric, in terms of which a Levi--Civita
connection can be defined. On the other hand, as is well known,
torsion and curvature are properties of a connection~\cite{livro},
and many different connections can be defined on the same space.
Therefore, in the specific case of a tetrad theory, we can say 
that the presence
of a nontrivial tetrad field in the gauge theory induces both, a
teleparallel and a riemannian structures in spacetime. The first is
related to the Cartan connection, a connection presenting torsion,
but no curvature. The second is related to the Levi--Civita
connection, a connection presenting curvature, but no torsion.
Then, owing to the universality of the gravitational interaction,
it turns out possible to link these geometrical structures with
gravitation. However, despite
the simultaneous presence of these two geometrical
structures, we will show in this paper that, in agreement with the 
equivalence alluded to above, the description of the gravitational 
interaction requires only one of the above structures. In other 
words, the gravitational interaction can be described alternatively 
in terms of magnitudes related to the teleparallel or to the riemannian 
structures induced in spacetime by the nontrivial tetrad field. 
Concerning the dynamics of the gravitational field, it has already 
been  shown~\cite{maluf} that this is in fact the case: the 
Hilbert--Einstein lagrangian of General Relativity, linear in the 
scalar curvature, is completely equivalent to the lagrangian of a 
translational gauge theory, quadratic in the torsion tensor. 
Similarly, we will show that the gravitational Lorentz force equation,
which describes the motion of a particle submitted to a gauge 
gravitational field in a flat spacetime, can be rewritten in terms of
magnitudes related to either, the teleparallel or the riemannian
geometry of spacetime. In the first case, the resulting equation is
not a geodesic but a force equation, which means that the
trajectories followed by scalar matter are not geodesics of the
induced Weitzenb\"ock spacetime. In the second case, the
gravitational Lorentz force becomes the geodesic equation of
General Relativity, which means that the trajectories followed by
scalar matter are geodesics of the induced Riemann spacetime. As
both descriptions are obtained from the same force equation, we
conclude that they are completely equivalent, which means 
essentially that scalar matter is able to feel anyone of the above 
spacetime geometries. 

\section{A Gauge Theory for the Translation Group}

We start by assuming spacetime to be a Minkowski space.
The second half of the greek alphabet ($\mu$, 
$\nu$, $\rho$,~$\cdots=1,2,3,4$) will be used to denote
indices related to this space. Its coordinates, 
therefore, will be denoted by $x^\mu$, and its metric
by $\eta_{\mu \nu}$.
At each point of spacetime, there is a tangent space
attached to it, given also by a Minkowski space, which will be 
the fiber of the corresponding tangent bundle.
We use the first half of the greek
alphabet ($\alpha$, $\beta$, $\gamma$,~$\cdots=1,2,3,4$) 
to denote indices related
to this space. Its coordinates, therefore, will be denoted by 
$x^\alpha$, and its metric by $\eta_{\alpha \beta}$. As
gauge transformations take place in this space, these will
also be the algebra indices of the gauge model.
The holonomic derivatives in these two spaces can be identified by
\be
\partial_\mu = \left(\partial_{\mu} x^{\alpha}\right) \;
\partial_\alpha \quad ; \quad \partial_\alpha =
\left(\partial_{\alpha} x^{\mu}\right) \;
\partial_\mu \; ,
\label{1}
\ee
where $\partial_{\mu} x^{\alpha}$ is a trivial holonomic 
tetrad, with $\partial_{\alpha} x^{\mu}$ its inverse.

A gauge transformation is defined as a local translation of the
fiber coordinates,
\be
x^{\prime \alpha} = x^{\alpha} + a^{\alpha}(x^\mu) \; ,
\label{traxf}
\ee
with $a^{\alpha}(x^\mu)$ the corresponding parameters. It can be
written in the form $x^{\prime} = U x$, where $U$ is an element of
the translation group. For an infinitesimal transformation,
\be
U = 1 + \delta a^\alpha P_\alpha \; ,
\label{uinfi}
\ee
with $\delta a^\alpha$ representing the infinitesimal parameters,
and $P_{\alpha} = {\partial}_{\alpha}$ standing for the generators
of infinitesimal translations, which satisfy
\be
\left[P_\alpha , P_\beta \right] = 0 \; .
\label{comu}
\ee
In terms of these generators, the infinitesimal version of 
transformation (\ref{traxf}) becomes
\be
\delta x^\alpha = \delta a^\beta P_\beta \; x^\alpha \; .
\label{traxg}
\ee

Let us consider now a general source field $\Phi(x^\mu)$. 
Its gauge transformation does not depend on the spin character, 
and is given by
\be
\Phi^{\prime}(x^\mu) = U \, \Phi(x^\mu) \; .
\label{gauphi}
\ee
The corresponding infinitesimal transformation, therefore, is
\be 
\delta \Phi = \delta a^\alpha P_\alpha \Phi \; ,
\label{trafi}
\ee
with $\delta \Phi$ standing for the functional change at the same
$x^\mu$, which is the relevant transformation for gauge theories.
It is important to remark that the translation generators are able
to act on any source field through their arguments because of
identification (\ref{1}). 

In order to define the gauge covariant derivative of
$\Phi(x^\mu)$, we must first introduce the gauge potentials of the
model, which will be denoted by 
\be
B_\mu = B^{\alpha}{}_{\mu} P_{\alpha} \; .
\label{b}
\ee
Using the general definition of gauge covariant derivatives,
\be
D_\mu = \partial_\mu + c^{-2} B^{\alpha}{}_{\mu} \,
\frac{\delta }{\delta a^\alpha} \; ,
\label{cova}
\ee
where the velocity of light $c$ was introduced for dimensional
reasons, the covariant derivative of $\Phi(x^\mu)$ turns out to be
\be
D_\mu \, \Phi = \partial_\mu \, \Phi + c^{-2} B^{\alpha}{}_{\mu} \;
P_\alpha \, \Phi \; .
\label{dfi1}
\ee
As the generators are derivatives which act on the fields through
their arguments, every source field in Nature will respond to
their action, and consequently will couple to the gauge potentials.
In other words, every source field in Nature will feel gravitation
the same way. This is the origin of the concept of {\em
universality} according to this model. 

{}From the covariance requirement for $D_\mu$, we can get the gauge
transformation of the potentials:
\be
B^{\prime}_{\mu} = U B_\mu U^{-1} + c^{2} U \partial_\mu U^{-1} \; .
\label{trab}
\ee
The corresponding infinitesimal transformation is
\be
B^{\prime \alpha}{}_\mu = B^{\alpha}{}_\mu - c^{2} \partial_\mu 
\delta a^\alpha \; .
\label{trabi}
\ee
The field strength $F^{\alpha}{}_{\mu \nu}$ is defined as the
covariant derivative of the gauge potential $B^\alpha{}_\mu$, and
analogously to the $U(1)$ electromagnetic gauge theory, it reads
\be
F^{\alpha}{}_{\mu \nu} = \partial_\mu B^\alpha{}_\nu - \partial_\nu
B^\alpha{}_\mu \; .
\label{fb}
\ee
Notice that, as expected for an abelian theory, $F^{\alpha}{}_{\mu
\nu}$ is invariant under a gauge transformation. Consequently, its
ordinary and gauge covariant derivatives coincide.

The dynamics of the gauge fields, as usual, can be obtained from a
lagrangian  quadratic in the field strength,
\be
{\cal L} = \frac{1}{16 \pi G} \left[ \frac{1}{4} \, 
F^{\alpha}{}_{\mu \nu} \, F^{\beta}{}_{\theta \rho} \, \eta^{\mu
\theta} N_{\alpha \beta}{}^{\nu \rho} \right] \; ,
\label{lm}
\ee
where $G$ is the gravitational constant, and
\be
N_{\alpha \beta}{}^{\nu \rho} = \eta_{\alpha \beta} \, \eta^{\nu
\rho} \; .
\label{n}
\ee 

\section{Gravitational Lorentz Force}

Let us now consider the motion of a particle of mass $m$ in a
gravitational field described by a translation gauge theory. As
spacetime is a Minkowski space, we can rely on an analogy with the
electromagnetic case to obtain  the corresponding gravitational
interaction. Thus, analogously to what  occurs in
electrodynamics~\cite{landau}, we assume the interaction of the
particle with the gravitational field to be described by the action
\be
c^{-2} \int_a^b B^\alpha{}_\mu \; p_\alpha \; dx^\mu \; ,
\label{intact}
\ee
with the integration taken along the world line of the particle.
In this expression, $p_\alpha$ is the Noether conserved charge
under the transformations of the gauge group~\cite{wong,drechsler}. 
In other words,
$p_\alpha$ is the four--momentum
$$
p_\alpha = m \, c \, u_\alpha
$$
of the particle, with $u_\alpha$ its four--velocity. Therefore, the
complete action for a particle in a gravitational field is given by
\be
S = \int_a^b L \, ds \equiv \int_a^b \left[- m \; c \; \sqrt{-
u^2} + \frac{m}{c} \; B^\alpha{}_\mu \; u_\alpha \; u^\mu \right]
ds \; ,
\label{totact}
\ee
with $ds=(\eta_{\mu \nu} dx^\mu \, dx^\nu)^{1/2}$ the Minkowski
interval,  and $u^2=\eta_{\mu \nu} u^\mu \, u^\nu$. Notice that, in
writing this action, we have already assumed the equality between
inertial and gravitational masses, as stated by the (weak)
equivalence principle~\cite{weinberg}. 

Next, we make use of the Euler--Lagrange equation
to obtain the equations of motion. The result is
\be
m c \left[ \frac{d u_\mu}{d s} + c^{-2} B^\alpha{}_\mu \frac{d
u_\alpha} {ds} \right] = \frac{m}{c} F^\alpha{}_{\mu \nu} \;
u_\alpha \; u^\nu \; .
\label{lore15}
\ee
Using the relation
$$
\frac{d u_\mu}{d s} = 
\left( \frac{\partial x^\alpha}{\partial x^\mu} \right) \; \frac{d
u_\alpha} {d s} \; ,
$$
we get finally
\be
\left(\partial_\mu x^{\alpha} + c^{-2} B^{\alpha}{}_{\mu} \right)
\; \frac{d u_\alpha}{d s} = c^{-2} \;
F^\alpha{}_{\mu \nu} \; u_\alpha \; u^\nu \; .
\label{lore1}
\ee 
This is the gravitational analog of the Lorentz
force. Its solution determines the
trajectory of the particle in a flat spacetime. It is interesting
to notice that, while in the electromagnetic case the particle
four--acceleration is proportional to $e/m$, with $e$ its  electric
charge, in the gravitational case the mass disappears from the
equation of motion. This is, of course, a consequence of the
assumed equivalence between the gravitational and inertial masses.

\section{The Induced Spacetime Geometry}

Up to this point, spacetime has been considered to be the Minkowski
space. However, as we are going to see, the presence of a nontrivial
tetrad field in the translation gauge theory induces further
structures in spacetime. Owing to the universality of the gravitational
interaction, it will then be possible to relate these structures to
the presence of gravitation. 

By using Eq.(\ref{1}), the covariant derivative (\ref{dfi1}) of a
general  source field can be rewritten in the form
\be
D_\mu \, \Phi = h^{\alpha}{}_{\mu} \; \partial_{\alpha} \, \Phi \; ,
\label{dfi2}
\ee
where
\be 
h^{\alpha}{}_{\mu} = \partial_{\mu} x^{\alpha} + c^{-2}
B^{\alpha}{}_{\mu} \equiv D_\mu \, x^\alpha
\label{h} 
\ee
is a tetrad field. Notice that the gravitational field appears as the
nontrivial part of the tetrad~\cite{kibble}. Making use
of Eqs.(\ref{traxg}) and (\ref{trabi}),  it is easy to see that, as in fact it
should be, the tetrad is gauge invariant: 
$$h^{\prime \alpha}{}_{\mu} = h^{\alpha}{}_{\mu} \; .$$
Its gauge covariant derivative, therefore, turns out to be the
same as  the ordinary derivative.

The expression for the gauge covariant derivative operator of
source fields,
\be
D_\mu = h^\alpha{}_\mu \; \partial_\alpha \; ,
\label{nbasis}
\ee
is actually the definition of a non--holonomous basis. In the
absence of gravitation, $h^\alpha{}_\mu$ becomes trivial, and it
reduces to the coordinate basis $\partial_\mu$ appearing in
Eq.(\ref{1}). This new basis, induced by the presence of the
gravitational field, satisfy the commutation relation
\be
\left[D_\mu , D_\nu \right] = c^{-2} \; F^\alpha{}_{\mu \nu} \;
P_\alpha \; .
\label{comuta}
\ee
Therefore, as usual in gauge theories, the commutator of covariant 
derivative operators yields the field strength $F^\alpha{}_{\mu
\nu}$.  There is a difference,  though: in contrast to the usual
gauge theories, the field strength here will be directly related
to the spacetime geometry. 

In fact, due to the presence of a tetrad field, there always
exists a  naturally defined linear Cartan connection~\cite{livro}
\be
\Gamma^{\rho}{}_{\mu \nu} = h_\alpha{}^\rho \, \partial_\nu
h^\alpha{}_\mu \; ,
\label{cone}
\ee
which is a connection presenting torsion, but no curvature.
As a consequence, the field strength $F^{\alpha}{}_{\mu \nu}$ can
be written  in the form
\be
F^{\alpha}{}_{\mu \nu} = c^{2} \; h^\alpha{}_\rho \; T^\rho{}_{\mu
\nu} \; ,
\label{ft}
\ee
where
\be
T^\rho{}_{\mu \nu} = \Gamma^{\rho}{}_{\nu \mu} -
\Gamma^{\rho}{}_{\mu \nu} \; 
\label{tor2}
\ee
is the torsion induced in spacetime by the presence of the 
gravitational field. Moreover, the commutation relation 
(\ref{comuta}) acquires the form
\be
\left[D_\mu , D_\nu \right] = T^\rho{}_{\mu \nu} \; D_\rho \; ,
\label{comuta2}
\ee
indicating that torsion plays also the role of the non--holonomy
of the  gauge covariant derivative.

The presence of a Cartan connection allows the introduction of a
spacetime covariant derivative which, acting for example on a
spacetime covariant vector $V_\mu$, reads
\be
{\nabla}_\nu \; V_\mu = \partial_\nu V_\mu -
\Gamma^{\theta}{}_{\mu \nu} \; V_\theta \; .
\label{telecode}
\ee
{}From this definition, one can easily see that, as a consequence
of  Eq.(\ref{cone}),
\be
{\nabla}_\nu \; h^{\alpha}{}_{\mu} =
\partial_\nu h^{\alpha}{}_{\mu} - h^{\alpha}{}_{\rho} \, 
\Gamma^{\rho}{}_{\mu \nu} \equiv 0 \; .
\label{weitz}
\ee
This is the condition of absolute parallelism, which implies that
the spacetime underlying a translational gauge theory is naturally
endowed  with a teleparallel structure. In other words, a
Weitzenb\"ock's  four--dimensional manifold is always present when
considering  a gauge theory for the translation group~\cite{hayshi}.

Besides the teleparallel structure, the presence of a nontrivial
tetrad induces also a riemannian structure in spacetime. As the
tetrad satisfies
\be
h^{\alpha}{}_{\mu} \; h_{\alpha}{}^{\nu} = \delta_{\mu}{}^{\nu}
\quad ; \quad h^{\alpha}{}_{\mu} \; h_{\beta}{}^{\mu} =
\delta^{\alpha}{}_{\beta} \; ,
\label{orto}
\ee
if algebra indices are raised and lowered with the
lorentzian metric $\eta^{\alpha \beta}$, tensor indices will
necessarily be raised and
lowered with the riemannian metric
\be
g_{\mu \nu} = \eta_{\alpha \beta} \; h^\alpha{}_\mu \;
h^\beta{}_\nu \; .
\label{gmn}
\ee
Accordingly, a linear metric connection can be introduced,
\be
{\stackrel{\circ}{\Gamma}}{}^{\theta}{}_{\mu \nu} = \frac{1}{2} 
g^{\theta \rho} \left[ \partial_{\mu} g_{\rho \nu} +
\partial_{\nu}  g_{\rho \mu} - \partial_{\rho} g_{\mu \nu} \right]
\; ,
\label{lc}
\ee
which is a connection presenting curvature, but no torsion. Its
curvature
\be
{\stackrel{\circ}{R}}{}^{\theta}{}_{\rho \mu \nu} = \partial_\mu
{\stackrel{\circ}{\Gamma}}{}^{\theta}{}_{\rho \nu} +
{\stackrel{\circ}{\Gamma}}{}^{\theta}{}_{\sigma \mu}
\; {\stackrel{\circ}{\Gamma}}{}^{\sigma}{}_{\rho \nu} - (\mu
\leftrightarrow \nu)
\; .
\label{rbola}
\ee
represents the curvature induced in spacetime by the presence of
the gravitational field. The connection ${\stackrel{\circ}{\Gamma}}
{}^{\theta}{}_{\mu \nu}$, actually the Levi--Civita connection of 
$g_{\mu \nu}$, allows the introduction of another spacetime 
covariant derivative which, acting for example 
on a spacetime covariant vector $V_\mu$, reads:
\be
{\stackrel{\circ}{\nabla}}{}_\nu \; V_\mu = \partial_\nu V_\mu - 
{\stackrel{\circ}{\Gamma}}{}^{\theta}{}_{\mu \nu} \; V_{\theta} \; .
\label{ricode}
\ee
As can be easily verified, both connections 
$\Gamma{}^{\theta}{}_{\mu \nu}$ and
${\stackrel{\circ}{\Gamma}}{}^{\theta}{}_{\mu \nu}$
preserve the metric:
$$
{\stackrel{\circ}{\nabla}}{}_\nu \; g_{\rho \mu} =
{\nabla}{}_\nu \; g_{\rho \mu} = 0 \; .
$$

Substituting now $g_{\mu \nu}$ into
${\stackrel{\circ}{\Gamma}}{}^{\theta}{}_{\mu \nu}$, we obtain
\be
{\Gamma}^{\theta}{}_{\mu \nu} =
{\stackrel{\circ}{\Gamma}}{}^{\theta} {}_{\mu \nu} +
{K}^{\theta}{}_{\mu \nu} \; ,
\label{decompo}
\ee
where
\be
{K}^{\theta}{}_{\mu \nu} = \frac{1}{2} \left[
T_{\mu}{}^{\theta}{}_{\nu} + T_{\nu}{}^{\theta}{}_{\mu} -
T^{\theta}{}_{\mu \nu} \right]
\label{conto}
\ee
is the contorsion tensor. Notice that the curvature of the Cartan 
connection vanishes identically:
\be
{R}^{\theta}{}_{\rho \mu \nu} = \partial_\mu
{\Gamma}^{\theta}{}_{\rho \nu} + {\Gamma}^{\theta}{}_{\sigma
\mu} \; {\Gamma}^{\sigma}{}_{\rho \nu} - (\mu  \leftrightarrow \nu)
\equiv 0 \; .
\label{r}
\ee
Substituting ${\Gamma}^{\theta}{}_{\mu \nu}$ from
Eq.(\ref{decompo}), we get
\be
{R}^{\theta}{}_{\rho \mu \nu} =
{\stackrel{\circ}{R}}{}^{\theta}{}_ {\rho \mu \nu} + \left(D K
\right)^{\theta}{}_{\rho \mu \nu} - {K}^{\theta}{}_{\sigma \mu}
\; {K}^{\sigma}{}_{\rho \nu} + {K}^{\theta}{}_{\sigma \nu} \;
{K}^{\sigma}{}_{\rho \mu} \equiv 0 \; ,
\label{eq7}
\ee
where
\be
\left(D K \right)^{\theta}{}_{\rho \mu \nu} =
\partial_\mu {K}^{\theta}{}_{\rho \nu} +
{\Gamma}{}^{\theta}{}_{\sigma \mu}
\; {K}^{\sigma}{}_{\rho \nu} + 
{\Gamma}{}^{\sigma}{}_{\rho \nu} \; {K}^{\theta}{}_
{\sigma \mu} - (\mu \leftrightarrow \nu) \; .
\ee

{}From these considerations, we conclude that the presence of a
nontrivial tetrad field in the translational gauge theory induces
both, a teleparallel and a riemannian structure in spacetime. These
structures together completely characterize the induced spacetime
geometry. Moreover, we see from Eq.(\ref{eq7}) that the Riemann
curvature tensor ${\stackrel{\circ}{R}}{}^{\theta}{}_{\rho \mu
\nu}$ induced in spacetime is such that it compensates exactly the
contribution to the curvature coming from the teleparallel
structure,  yielding a identically zero {\it total} curvature
tensor.  As we are going to see, however, the description of the
gravitational interaction requires only one of the above
structures. In other words, the gravitational interaction can be
described alternatively in terms of magnitudes related to the
teleparallel or to the riemannian geometry. 

\section{The Dynamics of the Gauge Fields}

In the induced spacetime, the gauge field lagrangian (\ref{lm}) is
written as
\be
{\cal L} = \frac{h}{16 \pi G} \left[ \frac{1}{4} \, 
F^{\alpha}{}_{\mu \nu} \, F^{\beta}{}_{\theta \rho} \, g^{\mu
\theta}  N_{\alpha \beta}{}^{\nu \rho} \right] \; ,
\label{lg}
\ee
where $h=\det (h^{\alpha}{}_{\mu})$ is the jacobian of the
transformation (\ref{nbasis}). Due to the presence of the tetrad
field, however, algebra and spacetime indices can now be  changed
into each other, and consequently appear mixed up in the
lagrangian. This means that
$$
N_{\alpha \beta}{}^{\nu \rho} = \eta_{\alpha \beta} \, g^{\nu
\rho} \equiv \eta_{\alpha \beta} \; h_{\gamma}{}^{\nu} \;
h^{\gamma \rho}
$$
must now include all cyclic permutations of $\alpha, \beta,
\gamma$.  A simple calculation shows that
\be
N_{\alpha \beta}{}^{\nu \rho} = 
\eta_{\alpha \beta} \; h_{\gamma}{}^{\nu} \; h^{\gamma \rho}
+ 2 \, h_{\alpha}{}^{\rho} \; h_{\beta}{}^{\nu}
- 4 \, h_{\alpha}{}^{\nu} \; h_{\beta}{}^{\rho} \; .
\label{ncy}
\ee
Substituting in (\ref{lg}), the gauge field lagrangian turns out
to be
\be
{\cal L} = \frac{h}{16 \pi G} F^{\alpha}{}_{\mu \nu} \,
F^{\beta}{}_{\theta \rho} \, g^{\mu \theta} \left[\frac{1}{4} \, 
h_\delta{}^\nu \, h^{\delta \rho} \, \eta_{\alpha \beta} +
\frac{1}{2} \, h_\alpha{}^\rho \, h_\beta{}^\nu - h_\alpha{}^\nu \,
h_\beta{}^\rho 
\right] \; .
\label{lagr1}
\ee
Then, by using Eq.(\ref{ft}), it becomes
\be
{\cal L} = \frac{h c^4}{16 \pi G} \left[\frac{1}{4} \;
T^\rho{}_{\mu \nu} \; T_\rho{}^{\mu \nu} + \frac{1}{2} \;
T^\rho{}_{\mu \nu} \; T^{\nu \mu}{}_\rho - T_{\rho
\mu}{}^{\rho} \; T^{\nu \mu}{}_\nu \right] \; .
\label{lagr2}
\ee

In order to obtain the vacuum field equations, it is convenient to 
rewrite it as~\cite{maluf}
\be
{\cal L} = \frac{h c^4}{16 \pi G} \; S_{\rho}{}^{\mu \nu} \;
T^{\rho}{}_{\mu \nu} \; ,
\ee
where
\be
S_{\rho}{}^{\mu \nu} = \frac{1}{4} \; \left(T_{\rho}{}^{\mu \nu} +
T^{\mu}{}_{\rho}{}^{\nu} - T^{\nu}{}_{\rho}{}^{\mu}\right) - \frac{1}{2} \;
\left(\delta_{\rho}{}^{\nu} \; T_{\theta}{}^{\mu \theta} -
\delta_{\rho}{}^{\mu} \; T_{\theta}{}^{\nu \theta} \right) \; .
\ee
By performing variations in ${\cal L}$ with respect to
$B^\alpha{}_\mu$, we get the equation
\be
\partial_\nu \; S_{\rho}{}^{\mu \nu} - 
\frac{4 \pi G}{c^4} \; t_{\rho}{}^{\mu} = 0 \; ,
\label{ym1}
\ee
where
\begin{eqnarray}
t_{\rho}{}^{\mu} &=& \frac{c^4}{16 \pi G} \; \Big[4 S_{\rho}{}^{\mu \theta}
\; \Gamma^{\nu}{}_{\nu \theta} + 2 S^{\mu \nu \theta} \;
T_{\rho \nu \theta} + T_{\sigma}{}^{\theta \mu} \; T^{\sigma}{}_{\theta \rho}
\nonumber \\
&+& T^{\theta}{}_{\sigma}{}^{\mu} \; T^{\sigma}{}_{\theta \rho}
- 2 T^{\theta \mu}{}_{\theta} \; T^{\theta}{}_{\rho \theta}
+ \frac{1}{2} \, T^{\mu \theta}{}_{\sigma} \; T_{\rho \theta}{}^{\sigma}
\Big] - \delta_{\rho}{}^{\mu} \; h^{-1} {\cal L}
\label{tem1}
\end{eqnarray}
is the gauge field self--current, which in this case is the
energy--momentum (pseudo) tensor of the gravitational field.
Notice that,  despite being an abelian gauge theory, this equation
presents a nonvanishing self--current which makes  it quite similar
to the Yang--Mills equations. However, in the present case, the
self--current has a different nature as, in contrast to the
standard Yang--Mills theories~\cite{ramond}, it can not be written
as a commutator since the gauge group here is abelian.

On the other hand, as we have already seen, the presence of a 
nontrivial tetrad field in the gauge theory induces also a
riemannian structure in spacetime. We consider then, in the
induced Riemann spacetime, the Hilbert--Einstein lagrangian of
General Relativity
\be
{\cal L} = \frac{c^4 h}{16 \pi G} \; {\stackrel{\circ}{R}} \; .
\label{ehl}
\ee
Substituting ${\stackrel{\circ}{R}}$ as obtained from
Eq.(\ref{eq7}), it can be rewritten in terms of the Cartan
connection only. Up to divergences, we get
\be
{\cal L} = \frac{h c^4}{16 \pi G} \; \left[\frac{1}{4} \;
T^\rho{}_{\mu \nu} \; T_\rho{}^{\mu \nu} + \frac{1}{2} \;
T^\rho{}_{\mu \nu} \; T^{\nu \mu} {}_\rho - T_{\rho \mu}{}^{\rho}
\; T^{\nu \mu}{}_\nu \right] \; ,
\label{lagr3}
\ee
which is exactly the lagrangian (\ref{lagr2}) of the translational
gauge  theory. We have in this way recovered the well known
result~\cite{maluf}  which says that  the translational gauge
theory, with a lagrangian quadratic in the torsion field, is
completely equivalent to General Relativity, with its usual
lagrangian linear in the  scalar curvature. As a consequence of 
the this equivalence,  the field equation (\ref{ym1}) of the
translational gauge theory must also be equivalent to the vacuum
Einstein's equation. In fact, through a tedious but straightforward
calculation, that  equation can be reduced to
\be
h \left({\stackrel{\circ}{R}}{}_{\rho \theta} - \frac{1}{2} \; 
g_{\rho \theta} \; {\stackrel{\circ}{R}}{} \right) = 0
\; .
\ee

\section{Gravitation as a Manifestation of Torsion: Force Equation}

Let us return to the gravitational force equation (\ref{lore1}),
which assumes now the form
\be
h^{\alpha}{}_{\mu} \; \frac{d u_\alpha}{d s} = c^{-2} \;
F^\alpha{}_{\mu \nu} \; u_\alpha \; u^\nu \; .
\label{lore5}
\ee
As we are going to see,
there are two different ways of interpreting this force equation.
In fact, it can be rewritten alternatively in terms of
magnitudes related to the Weitzenb\"ock or to the Riemann
spacetime,  giving rise respectively to the teleparallel and the
metric description of gravitation.

We start with the first alternative, which corresponds to
transform  algebra into tensor indices in such a way to get
the force equation (\ref{lore5}) written in terms of the Cartan
connection only. By using Eq.(\ref{ft}), as well as the relations
\be
h^{\alpha}{}_{\mu} \frac{d u_{\alpha}}{d s} = \omega_\mu \equiv
\frac{d u_{\mu}}{d s} - \Gamma_{\theta \mu \nu} u^{\theta}
\, u^{\nu} \quad ; \quad  u_{\alpha} = h_{\alpha \theta} \; 
u^{\theta} \; ,
\ee
where $\omega_\mu$ is the particle four acceleration in the induced 
spacetime, it reduces to
\be
\frac{d u_\mu}{d s} - \Gamma_{\theta \mu \nu} \; u^\theta \;
u^\nu =  T_{\theta \mu \nu} \; u^\theta \; u^\nu \; .
\label{geode}
\ee
The left--hand side of this equation is the Cartan covariant
derivative of $u_\mu$ along the world line of the particle. The
presence of the torsion tensor on its right--hand side, which plays
the role of an external force, implies that spinless particles do
not follow geodesics in the induced  Weitzenb\"ock spacetime.
Substituting Eq.(\ref{tor2}), it becomes
\be
\frac{d u_\mu}{d s} - \Gamma_{\theta \nu \mu} \; u^\theta \;
u^\nu = 0 \; .
\label{geodeflat}
\ee
Notice that, as $\Gamma_{\theta \nu \mu}$ is not symmetric in the
last two indices, this is in fact not a geodesic equation.
Actually, it is a force equation describing the interaction of a
spinless particle with the gravitational field. According to this
description, the only effect of the gravitational field  is to
induce a {\em torsion} in spacetime, which will then be the 
responsible for determining the trajectory of the particle.

\section{Gravitation as a Manifestation of Curvature: Geodesic
Equation}

Again, we transform algebra into spacetime indices, but now in
such a way  to get the force equation (\ref{lore5}) written in
terms of the Levi--Civita connection only. 
Following the same steps used earlier, we get
\be
\frac{d u_\mu}{d s} - \Gamma_{\theta \mu \nu} \; u^\theta \; u^\nu = 
T_{\theta \mu \nu} \; u^\theta \; u^\nu \; .
\label{geode5}
\ee 
Then, by taking into account the symmetry of $u^\theta \; u^\nu$
under the exchange $(\theta \leftrightarrow \nu)$, we can rewrite
it as
\be
\frac{d u_\mu}{d s} - \Gamma_{\theta \mu \nu} \; u^\theta \;
u^\nu =  K_{\mu \theta \nu} \; u^\theta \; u^\nu \; .
\label{geode2}
\ee
Noticing that $K_{\mu \theta \nu}$ is skew--symmetric in the
first two indices, and using Eq.(\ref{decompo}) to express
$(K_{\theta \mu \nu} - \Gamma_{\theta \mu \nu})$,
Eq.(\ref{geode2}) becomes
\be
\frac{d u_\mu}{d s} - {\stackrel{\circ}{\Gamma}}{}_{\theta \mu
\nu} \;  u^\theta \; u^\nu = 0 \; .
\label{geo2}
\ee
This is precisely the geodesic equation of General Relativity, 
which means that the trajectories followed by spinless particles 
are geodesics of the induced Riemann spacetime. 
According to this description, therefore, the only effect of the 
gravitational field is to induce a {\em curvature} in spacetime, 
which will then be the responsible for determining the
trajectory of the particle.

\section{Conclusions}

In the context of a gauge theory for the translation group, we
have  succeeded in obtaining a gravitational analog of the Lorentz
force equation. This force equation determines the trajectory to be
followed by a spinless particle submitted to a gauge gravitational
field, in a flat Minkowski spacetime.
According to this approach, the trajectory of the particle is
described in the very same way the Lorentz force describes the
trajectory of a charged particle in the presence of an
electromagnetic field. This force equation, however, can be rewritten 
in terms of magnitudes related to either, the telaparallel or the 
riemannian structures induced in spacetime by the presence of 
gravitation, which is represented by a nontrivial tetrad field.

The first alternative corresponds to consider the zero--curvature
Cartan  connection $\Gamma{}_{\theta \nu \mu}$ defined on
spacetime. Its torsion  will be the only manifestation of the
gravitational field in this case. In terms of this connection, the
force equation (\ref{lore1}) becomes
\be
\frac{d u_\mu}{d s} - \Gamma_{\theta \nu \mu} \; u^\theta \;
u^\nu = 0 \; .
\label{geodeflat1}
\ee
It is important to remark that, as $\Gamma_{\theta \nu \mu}$ is 
not symmetric in the last two indices, this is not a
geodesic equation, which means that the trajectories followed by
spinless particles are not geodesics of the  induced Weitzenb\"ock
spacetime. In a locally inertial coordinate system, the Cartan
connection $\Gamma_{\theta \nu \mu}$ becomes skew--symmetric  in
the first two indices, being this property the teleparallel 
version of  the normal coordinate condition $\partial_\mu
g_{\theta \nu} = 0$ of General Relativity.
In this coordinate system, therefore, owing to the symmetry of 
$u^\theta \; u^\nu$, the force equation (\ref{geodeflat1})
becomes the equation of motion of a free particle. This is the
teleparallel version of the (strong) equivalence principle.

The second alternative corresponds to consider the torsionless
Levi--Civita connection ${\stackrel{\circ}{\Gamma}}{}_{\theta \mu
\nu}$ defined on spacetime. Its curvature will be the only
manifestation of the gravitational field now. In this case, the
force equation (\ref{lore1})  is reduced to
\be
\frac{d u_\mu}{d s} - {\stackrel{\circ}{\Gamma}}{}_{\theta \mu
\nu} \;  u^\theta \; u^\nu = 0 \; .
\label{geo3}
\ee
This is the geodesic equation of General Relativity, which is an
equation  written in the underlying Riemann spacetime.
It corresponds mathematically to the vanishing of the Levi--Civita
covariant  derivative of $u_\mu$ along the world line of the
particle. In a locally inertial coordinate system, the first
derivative of the metric tensor vanishes, the Levi--Civita
connection vanishes as well, and the geodesic equation (\ref{geo3})
becomes the equation of motion of a free particle. This is the
usual version of the (strong) equivalence principle as formulated
in the General Theory of Relativity~\cite{weinberg}.

Notice the difference in the index contractions between the
connections and the four--velocities in equations
(\ref{geodeflat1}) and (\ref{geo3}). This difference is the
responsible  for the different characters of these equations: the
first is a force equation written in the underlying Weitzenb\"ock 
spacetime, and the second is a true geodesic equation written in
the induced Riemann spacetime. Furthermore, as can be easily 
verified, both  equations yield, for velocities sufficiently small, the 
usual Newtonian limit~\cite{weinberg}
$$
\phi \equiv - B_{0 0} = - \frac{G M}{r} \; ,
$$
with $\phi$ the Newton gravitational potential. 

It is interesting to remark that anyone of Eqs.(\ref{geodeflat1}) and
(\ref{geo3}) can be deduced from a variational principle with the action
$$
S = - \int_a^b m \; c \;  ds \; ,
$$
where $ds=(g_{\mu \nu} dx^\mu \, dx^\nu)^{1/2}$ is the spacetime
interval, and $g_{\mu \nu}= \eta_{\alpha \beta} h^{\alpha}{}_{\mu}
h^{\beta}{}_{\nu}$. In contrast to the Minkowskian action (\ref{totact}),
the interaction of the particle with the gravitational field is in this case
given by the presence of the metric tensor $g_{\mu \nu}$ in 
$ds$, or alternatively in $u^2=g_{\mu \nu} u^\mu \, u^\nu$ if one opts 
for using a lagrangian formalism.

Now, as both equations (\ref{geodeflat1}) and (\ref{geo3}) are
deduced from the same force equation (\ref{lore1}), they  must be
equivalent ways of describing the same physical trajectory. In
fact, it is easy to see that any one of them can be obtained from
the other by substituting the relation
\be
{\Gamma}_{\theta \mu \nu} = {\stackrel
{\circ}{\Gamma}}{}_{\theta \mu \nu} + {K}_{\theta \mu \nu} \; .
\label{decompo2}
\ee 
In General Relativity, the  presence of a gravitational field is
expressed by a torsionless  metric--connection, whose curvature
determines the intensity of the gravitational field, and
consequently the trajectories to be followed by spinless particles
under the influence of the gravitational field. On the other hand,
in the teleparallel description of gravitation, the presence of  a
gravitational field is expressed by a flat Cartan connection,
whose torsion is now the entity responsible for determining the
intensity of the gravitational field, and consequently the
trajectories to be followed by spinless particles under the
influence of the gravitational field. Thus, we can say that the
gravitational interaction can  be described {\em either}, in terms
of the curvature of spacetime, as is usually done in General
Relativity, or in terms of the torsion  of spacetime. Both
interpretations result completely equivalent in the sense that they
give  the same physical trajectory for a spinless particle in a
gravitational  field. Whether gravitation requires a curved or a
torsionned spacetime, therefore,  turns out to be a matter of
convention. Moreover, contrary to the old belief~\cite{nitsch2}
that only particles  with spin could detect the teleparallel
geometry, scalar matter being  able to feel the metric geometry
only, our results imply that scalar matter  is able to feel anyone
of these geometries.

\section*{Acknowledgments}

The authors would like to thank R. Aldrovandi for useful
discussions, and for a critical reading of the manuscript. They
would also like to thank CNPq--Brazil, for financial support.


\begin{thebibliography}{10}

\bibitem{moller}
C. M{\o}ller, K. Dan. Vidensk. Selsk. Mat. Fys. Skr.{\bf 1}, No.
10 (1961).

\bibitem{pelle}
C. Pellegrini and J. Plebanski, K. Dan. Vidensk. Selsk. Mat. Fys.
Skr. {\bf 2}, No. 2 (1962).

\bibitem{haya}
K. Hayashi and T. Nakano, Prog. Theor. Phys. {\bf 38}, 491 (1967).

\bibitem{hayshi}
K. Hayashi and T. Shirafuji, Phys. Rev. D {\bf 19}, 3524 (1979).

\bibitem{hehl}
F. W. Hehl, in {\em Cosmology and Gravitation}, ed. by P. G.
Bergmann and  V. de Sabbata (Plenum, New York, 1980).

\bibitem{kopc}
W. Kopczy\'nski, J. Phys. A {\bf 15}, 493 (1982).

\bibitem{azeredo}
R. de Azeredo Campos and C. G. Oliveira, Nuovo Cimento B {\bf 74},
83 (1983).

\bibitem{nitsch1}
F. M\"uller--Hissen and J. Nitsch, Gen. Rel. Grav. {\bf 17}, 747
(1985).

\bibitem{mielke}
E. W. Mielke, Ann. Phys. (NY) {\bf 219}, 78 (1992).

\bibitem{hene}
F. W. Hehl, J. D. McCrea, E. W. Mielke and Y. Ne'eman, Phys. Rep. 
{\bf 258}, 1 (1995).

\bibitem{weitz}
R. Weitzenb\"ock, {\em Invariantentheorie} (Noordhoff, Gronningen,
1923).

\bibitem{livro}
R. Aldrovandi and J. G. Pereira, {\em An Introduction to
Geometrical Physics} (World Scientific, Singapore, 1995).

\bibitem{maluf}
J. W. Maluf, J. Math. Phys. {\bf 35}, 335 (1994).

\bibitem{landau}
L. D. Landau and E. M. Lifshitz, {\em The Classical Theory of
Fields}  (Pergamon, Oxford, 1971).

\bibitem{wong}
S. K. Wong, Nuov. Cim. {\bf 65}, 689 (1970).

\bibitem{drechsler}
W. Drechsler, Phys. Lett. B {\bf 90}, 258 (1980).

\bibitem{weinberg}
See, for example, S. Weinberg, {\em Gravitation and Cosmology} 
(Wiley, New York, 1972).

\bibitem{kibble}
T. W. B. Kibble, J. Math. Phys. {\bf 2}, 212 (1961).

\bibitem{ramond}
See, for example, P. Ramond, {\em Field Theory: a Modern Primer}
(Benjamin/Cummings, Reading, 1981).

\bibitem{nitsch2}
J. Nitsch and F. W. Hehl, Phys. Lett. B {\bf 90}, 98 (1980).

\end{thebibliography}
\end{document}